\documentclass[a4paper,12pt]{article}

\usepackage{indentfirst}

\usepackage{amsfonts,amsmath,amssymb,bm,a4wide,graphicx,cite,makeidx,multicol}

\begin{document}

\title{Spin light of neutrino in matter:
a new type of electromagnetic radiation}

\author{Alexander Grigoriev
   \\
   \small {\it Skobeltsyn Institute of Nuclear Physics, Moscow State University, 119992 Moscow,  Russia }
   \\
        Andrey Lobanov , \
        Alexander Studenikin
   \\
   \small {\it Department of Theoretical Physics, Moscow State University, 119992 Moscow,  Russia }
   \\
   Alexei Ternov
   \\
   \small{\it Department of Theoretical Physics, Moscow Institute for Physics and Technology,}
   \\
   \small {\it 141700 Dolgoprudny, Russia }}

\date{}
\maketitle

\sloppy

\begin{abstract}
A short review of the properties of the spin light of neutrino
($SL\nu$) in matter, supplied with some historical notes on the
discussed subject, is given. It is shown that consideration of the
$SL\nu$ in matter in hep-ph/0605114 is based on erroneous
calculations which ignore the fact that the energy-momentum
conservation law can not be violated for this process. An attempt
to rename the $SL\nu$ in matter, undertaken in hep-ph/0606262, is
groundless.
\end{abstract}

In a series of our papers \cite{LobStudPLB03} -\cite{StuJPA06}, we
have proposed and studied in detail a new type of electromagnetic
radiation that can be emitted by a massive neutrino with nonzero
magnetic moment moving in background matter. We have termed this
radiation the \ ``spin light of neutrino" \ ($SL\nu$)
\cite{LobStudPLB03}. At first we have developed the
quasi-classical theory of this radiation on the basis of the
generalized Bargmann-Michel-Telegdi equation that we have derived
\cite{EgoLobStuPLB00LobStuPLB01}, \cite{DvoStuJHEP02} for
description of the neutrino spin evolution in the presence of
matter. As it was clear from the very beginning
\cite{LobStudPLB03}, the $SL\nu$ is a quantum phenomenon by its
nature. Therefore, we later on considered the $SL\nu$ on a solid
base of the modified Dirac equation for the neutrino wave function
in matter and elaborated \cite{StuTerPLB05} -\cite{StuJPA06} the
quantum theory of this radiation.

The main results of the performed studies \cite{LobStudPLB03}
-\cite{StuJPA06} of the $SL\nu$ in matter enable us to summarize
the properties of this process as follows \cite{GriStuTerPLB_05}:

1) a neutrino with nonzero mass and magnetic moment can emit spin
light when moving in dense matter;

2) in general,  $SL\nu$ in matter is due to the dependence of the neutrino dispersion relation in
matter on the neutrino helicity;

3) the $SL\nu$ radiation rate and  power depend on the neutrino magnetic moment and energy, and
also on the matter density;

4)  the matter density parameter, which depends on the type of
neutrino and matter composition,  can be negative; therefore the
types of initial and final neutrino (and antineutrino) states,
conversion between which effectively produces the $SL\nu$
radiation, are determined by the matter composition;

5) the $SL\nu$ in matter leads to the neutrino-spin polarization effect; depending on the type of
the initial neutrino (or antineutrino) and matter composition the negative-helicity relativistic
neutrino (the left-handed neutrino $\nu_{L}$) is converted to the positive-helicity neutrino (the
right-handed neutrino $\nu_{R}$) or vice versa;

6) the obtained expressions for the $SL\nu$ radiation rate and power exhibit non-trivial dependence
on the density of matter and on the initial neutrino energy; the $SL\nu$ radiation rate and power
are proportional to the neutrino magnetic moment squared which is, in general, a small value and
also on the neutrino energy, that is why the radiation discussed can be effectively produced only
in the case of ultra-relativistic neutrinos;

7) for a wide range of matter densities the radiation is beamed
along the neutrino momentum, however the actual shape of the
radiation spatial distribution may vary from projector-like to
cap-like, depending on the  neutrino momentum-to-mass ratio and
the matter density;

8) in a wide range of matter densities the $SL\nu$ radiation is
characterized by total circular polarization;

9) the emitted photon energy is also essentially dependent on the
neutrino energy and matter density; in particular, in the most
interesting for possible astrophysical and cosmology applications
case of ultra-high energy neutrinos, the average energy of the
$SL\nu$ photons is one third of the neutrino momentum.

Considering  the listed above properties of the $SL\nu$ in matter,
we argue that this radiation may be produced by high-energy
neutrinos propagating in different astrophysical and cosmological
environments.

Here we should like to mention that the considered $SL\nu$ is indeed a new type of electromagnetic
radiation of neutrino that can be emitted by neutrino and that has never been considered before. As
it was mentioned in our first paper on this subject \cite{LobStudPLB03}, the proposed mechanism of
radiation is totally different from many other known processes characterized by the same signature,
$\nu\rightarrow \nu + \gamma$, as one of the $SL\nu$, for instance:

 i) the photon radiation by massless neutrino
$(\nu_{i} \rightarrow \nu_{j} + \gamma,\; i=j)$ due to the vacuum
polarization loop diagram  in presence of an external magnetic
field \cite{GalNik72,Sko76};

ii) the photon radiation by massive neutrino with non-vanishing
magnetic moment in constant magnetic and electromagnetic wave
fields \cite{BorZhuTer88,Sko91};

iii) the Cherenkov radiation due to the non-vanishing neutrino
magnetic moment in homogeneous and infinitely extended medium
which is only possible if the speed of neutrino is larger than the
speed of light in medium \cite{Rad75,GriNeu93};

iv) the transition radiation due to non-vanishing neutrino
magnetic moment which would be produced when the neutrino crosses
the interface of two media with different refractive indices
\cite{SakKur94-95,GriNeu95};

v) the Cherenkov radiation by  massless neutrino due to its
induced charge in medium \cite{OliNiePal96};

vi) the Cherenkov radiation by massive and massless  neutrino in
magnetized medium \cite{MohSam96, IoaRaf97};

vii) the neutrino radiative decay $(\nu_{i} \rightarrow \nu_{j} + \gamma, \; i\not=j)$ in external
fields and medium or in vacuum \cite{GvoMikVas92,Sko95,ZhuEmiGri96,KacWun97,TerEmi03}.

Thus, that the proposed and studied in our papers \cite{LobStudPLB03} -\cite{StuJPA06} mechanism of
electromagnetic radiation generated by the neutrino magnetic moment which occurs due to electroweak
interaction of a neutrino  with the background environment has never been considered before. It
should be emphasized that the spin light of neutrino  can not be described as the Cherenkov
radiation, because the $SL\nu$ is produced even in the case when modification of the photon
dispersion relation by the environment can be neglected.

Note that possible influence of medium on emitted photons was also
discussed in our papers
\cite{GriStuTerPLB_05,GriStuTerCOSMION04_hep_ph0502231}. It was
shown \cite{GriStuTerPLB_05,GriStuTerCOSMION04_hep_ph0502231} that
for the most interesting case of high-energy neutrinos the matter
influence on the photon dispersion can be neglected. In addition,
here it should be noted that, as it is well known (see, for
instance, \cite{Ginz60} and \cite{Shef75}), plasma is transparent
for electromagnetic radiation on frequencies greater than the
plasmon frequency. In \cite{LobStudPLB03} -\cite{StuJPA06}, we
have shown from the energy-momentum conservation law that a
relativistic neutrino can emit the $SL\nu$ photons with
characteristic energy equals to a reasonable fraction of neutrino
energy. So that if neutrino energy much exceeds the plasmon
frequency then the plasma influence on photons can be neglected
\cite{GriStuTerPLB_05, GriStuTerCOSMION04_hep_ph0502231}.

The phenomenon of $SL\nu$ in matter has attracted attention of another authors. In the preprint
hep-ph/0605114 ( V 1 of May 10, 2006) \cite{KuznMikhhep_ph0605114}, it has been claimed that there
is \ ``no neutrino spin light because of photon dispersion in medium". In our remark
\cite{GriLobStuTer_hp0606011} (hep-ph/0606011, V 1 of June 1, 2006) to this statement, we have
explained why the mentioned above conclusion of \cite{KuznMikhhep_ph0605114} is wrong. As it
follows from the consideration undertaken in \cite{KuznMikhhep_ph0605114}, the mentioned above
false statement is based on uncorrect evaluation of the $SL\nu$ photon energy in which the authors
of \cite{KuznMikhhep_ph0605114} ignore the momentum conservation law. It is obvious that the
fundamental law of the energy-momentum conservation can not be violated in any process, including
the $SL\nu$ in matter.

The same authors have further continued their studies on the $SL\nu$ in matter. Recently, without
any reference to the first their preprint \cite{KuznMikhhep_ph0605114} and our remark
\cite{GriLobStuTer_hp0606011} the authors of \cite{KuznMikhhep_ph0605114} have issued the second
preprint hep-ph/0606262 ( V 1 of June 25, 2006) and published the exactly identical paper
\cite{KuznMikhhep_ph0606262_MPL06} entitled \ ``Plasma induced neutrino radiative decay instead of
neutrino spin light". In these new studies \cite{KuznMikhhep_ph0606262_MPL06}, the authors have
undertaken an attempt to consider, as it is written in \cite{KuznMikhhep_ph0606262_MPL06}, \ `` the
conversion of a neutrino with a magnetic moment... caused by the additional Wolfenstein energy
acquired by a left-handed neutrino in medium, with an accurate account of the photon dispersion in
medium". Obviously, it is the $SL\nu$ in matter that is considered in
\cite{KuznMikhhep_ph0606262_MPL06} with addition account for the photon dispersion in medium.

As it follows from the title of the paper
\cite{KuznMikhhep_ph0606262_MPL06}, the authors try to allege that
inclusion of the photon medium dispersion in the $SL\nu$ effect
sufficiently changes the nature of this phenomena, so that the new
title instead of the \ ``spin light of neutrino" is needed.
However, as it is clear from the studies of
\cite{KuznMikhhep_ph0606262_MPL06}, the authors deals with the
same phenomenon that was first introduced and studied in detail in
our papers \cite{LobStudPLB03} -\cite{StuJPA06} and then was also
discussed in their first preprint  \cite{KuznMikhhep_ph0605114},
where there have been no word said about the need to rename the
spin light of neutrino. The fact is that the \ ``new effect"
introduced in \cite{KuznMikhhep_ph0606262_MPL06} appears, as the
authors say themselves in \cite{KuznMikhhep_ph0606262_MPL06}, due
to 1) the neutrino magnetic moment interaction with photons, 2)
the additional energy acquired by neutrino in matter. This is just
exactly the same mechanism that one of the $SL\nu$ in matter
\cite{LobStudPLB03} -\cite{StuJPA06}.

We should like to mention that the introduced and used in our
papers \cite{LobStudPLB03} -\cite{StuJPA06} term  \ ``spin light
of neutrino" has a famous precursor of the same physics nature.
Indeed, the $SL\nu$ in matter is an example of a more general sort
of radiation previously termed ``spin light'' \cite{SL_theor},
which is just radiation of an intrinsic magnetic moment of an
electron associated with its spin. In the case of the
``synchrotron radiation'' (i.e. the radiation of a relativistic
charged particle in an external magnetic field), its dependence on
the electron spin orientation was studied both theoretically
\cite{SL_theor} and also experimentally in the Budker INP
(Novosibirsk) \cite{SL_exper}. As a result of these studies, it
has become clear that the synchrotron radiation can be considered
as if it consists of the radiation of an electron charge itself
and the one of an intrinsic magnetic moment of an electron. The
latter is just what was called the ``spin light''.

    In connection with the attempt \cite{KuznMikhhep_ph0606262_MPL06}
to rename the $SL\nu$ in matter to ``plasma induced neutrino
radiative decay", we should like to remind that the proposed in
\cite{KuznMikhhep_ph0606262_MPL06} term  \ ``decay" has been
already used for designation of various transitions between
different types of neutrinos with emission of photons (or
plasmons)
\cite{GvoMikVas92,Sko95,ZhuEmiGri96,KacWun97,IoaRaf97,TerEmi03},
the processes of quite different nature that one considered in
\cite{KuznMikhhep_ph0606262_MPL06}.

Although the obtained in \cite{KuznMikhhep_ph0606262_MPL06} final expression for the total width of
the considered process in the most interesting limiting case of ultra-high neutrino energies gives
exactly the result that have been obtained previously in our papers
\cite{GriStuTerCOSMION04_hep_ph0502231} -\cite{LobDAN05} (for some reason this fact is not pointed
out in \cite{KuznMikhhep_ph0606262_MPL06}), it is not clear how this result was obtained in
\cite{KuznMikhhep_ph0606262_MPL06}. The fact is that the studies of the process in
\cite{KuznMikhhep_ph0606262_MPL06} are based on the ill-defined expression for the process matrix
element squared. The expression for the matrix element squared given by formula (18) of
\cite{KuznMikhhep_ph0606262_MPL06} is not positively-defined. One may wonder how this result, as
the authors wrote in \cite{KuznMikhhep_ph0606262_MPL06} while explaining the derivation of formula
(18), ``can be obtained by the standard way".

One of the authors (A.S.) would like to thank Victor Matveev and Valery Rubakov
for the invitation to participate in the 14th International Seminar on High
Energy Physics "Quarks-2006", and also to all of the organizers for their kind
hospitality in Repino.


\vspace{1cm}





\begin{thebibliography}{99}



\bibitem{LobStudPLB03} A.Lobanov, A.Studenikin, Phys.Lett.B564
(2003), 27, hep-ph/0212393.

\bibitem{LobStudPLB04} A.Lobanov, A.Studenikin, Phys.Lett.B601 (2004) 171.

\bibitem{DvoGriStudIJMPD05} M.Dvornikov, A.Grigoriev, A.Studenikin,
Int.J.Mod.Phys.D14 (2005) 309.
\bibitem{DvoStuJHEP02}
 M.Dvornikov, A.Studenikin, JHEP 09 (2002) 016.
\bibitem{StuNPB05} A.Studenikin, Nucl.Phys.(Proc.Suppl.)B 143 (2005) 570.

\bibitem{StuTerPLB05} A.Studenikin, A.Ternov, Phys.Lett.B608 (2005) 107.
\bibitem{GriStuTerCOSMION04_hep_ph0502231} A.Grigoriev, A.Studenikin, A.Ternov,
Grav. \& Cosm. 11 (2005) 132.

\bibitem{LobPLB05} A.Lobanov, Phys.Lett.B 619 (2005) 136.

\bibitem{GriStuTerPLB_05} A.Grigoriev, A.Studenikin, A.Ternov,
Phys.Lett.B 622 (2005) 199.


\bibitem{LobDAN05} A.Lobanov, Dokl.Phys. 50 (2005) 286.


\bibitem{StuJPA06} A.Studenikin,
J.Phys.A: Math. Gen. 39 (2006) 6769.

\bibitem{EgoLobStuPLB00LobStuPLB01}
 A.Egorov, A.Lobanov, A.Studenikin, Phys.Lett.B491 (2000) 137;
 A.Lobanov, A.Studenikin, Phys.Lett.B515 (2001) 94.

\bibitem {GalNik72} D.V.Galtsov, N.S.Nikitina, Sov.Phys.JETP 35 (1972) 1047,
(Zh.Eksp.Teor.Fiz. 62 (1972) 2008).


\bibitem {Sko76} V.V.Skobelev, Sov.Phys.JETP 44 (1976) 660,
(Zh.Eksp.Teor.Fiz. 71 (1976) 1263).

\bibitem {BorZhuTer88} A.V.Borisov, V.Ch.Zhukovskii, A.I.Ternov,
Sov.Phys.J. 31 (1988) 228 (Izv.Vuzov.Fiz. No. 3 (1988) 64).

\bibitem {Sko91} V.V.Skobelev, Sov.Phys.JETP 73 (1991) 40,
(Zh.Eksp.Teor.Fiz. 100 (1991) 75).

\bibitem {Rad75} M.Radomski, Phys.Rev.D12 (1975) 2208.

\bibitem  {GriNeu93} W.Grimus, H.Neufeld, Phys.Lett.B315 (1993) 129.

\bibitem {SakKur94-95} M.Sakuda, Phys.Rev.Lett.72 (1994) 804, M.Sakuda,
Y.Kurihara, Phys.Rev.Lett.74 (1995) 1284

\bibitem {GriNeu95} W.Grimus, H.Neufeld, Phys.Lett.B344 (1995)
252.

\bibitem {OliNiePal96} J.C.D'Olivo, J.Nieves, P.B.Pal, Phys.Lett.B365 (1996)
178.

\bibitem {MohSam96} S.Mohanty, M.Samal, Phys.Rev.Lett.77 (1996).

\bibitem {IoaRaf97} A.Ioannisian, G.G.Raffelt, Phys.Rev.D55 (1997) 7038.

\bibitem {GvoMikVas92} A.A.Gvozdev, N.V.Mikheev, L.A.Vassilevskaya,
Phys.Lett.B289 (1992) 103.

\bibitem {Sko95} V.V.Skobelev, Sov.Phys.JETP 81 (1995) 1,(Zh.Eksp.Teor.Fiz. 108 (1995) 3).
\bibitem {ZhuEmiGri96} V.Ch.Zhukovskii, P.A.Eminov, A.E.Grigoruk,
Mod.Phys.Lett.A11 (1996) 3113 .
\bibitem {KacWun97} M.Kachelriess, G.Wunner, Phys.Lett.B390 (1997) 263.
\bibitem {TerEmi03} A.I.Ternov, P.A.Eminov, J.Phys.G29 (2003) 357.

\bibitem{Ginz60}V.L.Ginzburg, The Propagation of Electromagnetic Waves in Plasmas,
Oxford: Pergamon Press, 1964.

\bibitem{Shef75} J.Sheffield, Plasma scattering of electromagnetic radiation,
New York, San Francisco: Academic Press, 1975.

\bibitem{KuznMikhhep_ph0605114}
A.Kuznetsov, N.Mikheev, hep-ph/0605114, V 1 of May 10, 2006.
\bibitem{GriLobStuTer_hp0606011}
A.Grigoriev, A.Lobanov, A.Studenikin, A.Ternov, hep-ph/0606011, V 1 of June 1,
2006.
\bibitem{KuznMikhhep_ph0606262_MPL06}
A.Kuznetsov, N.Mikheev, hep-ph/0606262, V 1 of June 25, 2006;
Mod.Phys.Lett. A23 (2006) 1769.
\bibitem{SL_theor} I.M.Ternov, Sov.Phys.Usp. 38 (1995) 405;
V.A.Bordovitsyn, I.M.Ternov, V.G.Bagrov, Sov.Phys.Usp. 38 (1995)
1037.
\bibitem{SL_exper}
S. A. Belomestnykh, A. E. Bondar, M. N. Yegorychev, V. N. Zhilich,
  G. A. Kornyukhin, S. A. Nikitin, E. L. Saldin, A. N. Skrinsky,
  G. M. Tumaikin, Nucl. Instr. \& Methods in
  Physics Research A, {\bf 227}, 173 (1984).
\end{thebibliography}
\end{document}